# The unbiased Diffusion Monte Carlo: a versatile tool for two-electron systems confined in different geometries

Gaia Micca Longo[1,2], Carla Maria Coppola[1,3], Domenico Giordano[4], Savino Longo[1,2]

[1] Department of Chemistry – Università degli Studi di Bari Aldo Moro – Via Orabona 4 – 70125 Bari, Italy
[2] Istituto per la Scienza e Tecnologia dei Plasmi – Consiglio Nazionale delle Ricerche, Bari Section – Via Amendola 122/D – 70125 Bari, Italy
[3] Osservatorio Astrofisico di Arcetri, Largo Enrico Fermi 5 Firenze – ITALY
[4] European Space Agency – ESTEC (Retired) – The Netherlands

Email: *gaia.miccalongo@uniba.it, carla.coppola@uniba.it, dg.esa.retired@gmail.com, savino.longo@uniba.it*

**Abstract**
Computational codes based on the Diffusion Monte Carlo method can be used to determine the quantum state of two-electron systems confined by external potentials of various nature and geometry. In this work, we show how the application of this technique in its simplest form, that does not employ complex analytic guess functions, allows to obtain satisfactory results and, at the same time, to write programs that are readily adaptable from one type of confinement to another. This adaptability allows an easy exploration of the many possibilities in terms of both geometry and structure of the system. To illustrate these results, we present calculations in the case of two-electron hydrogen-based species ($H_2$ and $H_3^+$) and two different types of confinement, nanotube-like and octahedral crystal-field.

## 1. Introduction

The fundamental relevance and concrete applications of the confined quantum systems has been established since the early days of quantum physics [Michels, 1937; Sommerfeld, 1938; De Groot, 1946], up to the present day [Sabin, 2009; Sen, 2009; Sen, 2014; Ley-Koo, 2018]. Simple confined physical systems have a very long tradition in physics because of their importance as basic theoretical problems [Aquino, 2007; Burrows, 2006; Micca Longo, 2015a; Micca Longo, 2015b; Micca Longo, 2015c; Longo, 2018], and as prototypes to gain insights in semiconductor physics (quantum dots [Chuu, 1992; Porras-Montenegro, 1992], quantum wires and quantum wells [Harrison, 2005; Munjal, 2018]). Atoms imprisoned in zeolite traps, clusters and fullerene cages provide good examples of confined systems in atomic physics and inorganic chemistry [Belosludov, 2003; Connerade, 1999; Hernandez-Rojas, 1996; Soullard, 2004].
Many of these studies on the physical-chemical properties of confined systems focus on systems with a small number of electrons: for example, species with small molecules containing hydrogen, helium and lithium. Furthermore, confinement is produced by means of an infinite potential well with different geometries, in order to produce analogies with different real systems without considering their details.

In recent years, our group has developed a very simple and versatile methodology to study this kind of systems [Micca Longo, 2015a; Micca Longo, 2015b; Micca Longo, 2015c; Longo, 2018]. It is a modification of the Diffusion Monte Carlo (DMC) method: since the number of electrons is low, no variance reduction technique requesting a trial wavefunction is employed; instead, some efficient solutions to accelerate calculations and reduce noise are employed. Thanks to this method, it is possible to study different systems, with different orientations, and even quickly change the geometry of the potential well, from spherical to elliptical, from cylindrical to cubical, using Cartesian coordinates and implementing few modifications to the starting code. Our previous works [Micca Longo, 2015a; Micca Longo, 2015b; Micca Longo, 2015c; Longo, 2018] focused on the effects of various confinement geometries on the excited states of monoelectronic systems, such as H and $H_2^+$. In this paper, we show the application of the method to the ground state of two-electron systems ($H_2$ and $H_3^+$) with non-trivial confinement geometries.

An exhaustive report on the many-electron confined system can be found in the work by Jaskólski [Jaskólski, 1996]: different methods of analysis and description of spatial confinement effects are reviewed, with a particular attention to their importance in semiconductor structures.

Free $H_2$ is one of the simplest molecular systems, and its electronic properties represent a key-point in understanding larger molecular systems; therefore, its confinement has received much attention during the last decades. The molecular hydrogen confined in a rigid spheroidal box, with fixed and not fixed nuclear positions, was studied with variational calculations and quantum Monte Carlo techniques [LeSar, 1981; LeSar, 1983; Pang, 1994; Colín-Rodríguez, 2010; de Oliveira Batel, 2018; Doma, 2018], also with Monte Carlo methods beyond the Born-Oppenheimer approximation [Sarsa, 2013].

Many studies concerning the ground state and the potential energy surfaces of the $H_3^+$ can be found in the literature [Conroy, 1964; Röhse, 1994; Jin, 2000; Pavanello, 2009; Adamowicz, 2012; Mizus, 2018; to cite a few], but almost no information is available about the quantum confinement of this interesting molecular system. Quantum calculations based on diffusion for the study of the free $H_3^+$ in equilateral triangle configuration trace back to the works by Anderson [Anderson, 1975; Anderson, 1992].

In this paper, we present the first results concerning the confined $H_2$ inside a six-negative-charge cage and the collinear $H_3^+$ enclosed in a nanotube. All results are obtained by diffusion Monte Carlo method.

## 2. Method

Quantum Monte Carlo (QMC) is a class of computer algorithm that are able to simulate quantum systems and to compute the electronic ground state of atoms, molecules and solids.

Among several Quantum Monte Carlo methods available [Foulkes, 2001], the diffusion Monte Carlo (DMC) method offers the possibility to consider confining walls of any geometry while retaining the use of cartesian coordinates and to include any distribution of positive charge within the cavity. Basically, DMC is a stochastic projector method that makes use of the similarity between the imaginary time Schrödinger equation and a generalized diffusion equation, which can be solved using stochastic calculus and simulating a random walk. The diffusion process of numerous "walkers" in the phase space is simulated. A walker is a mobile mathematical point in a set $\{\boldsymbol{R}_i\}$ of phase coordinates *R* in a *3N*-dimensional space, where *N* is the number of *physical* electrons in the system, different from the number of walkers *M*, which is an integer numerical variable, $M \sim 10^3$. An evolution is performed by repeated application of the short-time diffusion-reaction propagator

$$G(\mathbf{R} \leftarrow \mathbf{R}', \tau) \approx (2\pi\tau)^{\frac{-3N}{2}} exp\left[\frac{-(\mathbf{R}-\mathbf{R}')^2}{2\tau}\right] exp\left[\frac{-\tau(V(\mathbf{R})-V(\mathbf{R}')-2E_T)}{2}\right] \quad (1)$$

where $\tau$ is a numerical step, and $E_T$ is the so-called energy offset that controls the walker total population. The kinetic energy term of the wave equation is connected with the diffusion process (the first exponential term of equation (1), including the normalizing constant), while the potential energy term (the second exponential term) controls the so called "birth-death algorithm", that is the destruction or the multiplication of walkers [Foulkes, 2001]. According to this second factor, a walker may randomly disappear, or duplicate itself, in a numerical time step $\tau$.

Each walker, say the *i*-th one, changes its state randomly at each calculation step: in the code, these coordinates are represented by vectors $x(i,j), j = 1, \dots, D$, so that even the change in dimensionality *D* is immediate. The micro propagator of equation (1) is the product of two factors: the first one

$$P_1 = (2\pi\tau)^{\frac{-3N}{2}} exp\left[\frac{-(\mathbf{R}-\mathbf{R}')^2}{2\tau}\right] \quad (2)$$

represents a diffusion process. To implement equation (2), we use a special technique: the central limit theorem of probability theory guarantees that, in the repeated application of random displacements, a Gaussian distribution like $P_1$ is obtained anyway. Consequently, our programs do not use more computationally expensive Gaussian random generators, but shift each coordinate according to a uniform probability distribution with the same variance as $P_1$:

$$x(i,j) \leftarrow x(i,j) + 3^{1/2}\tau(2\eta - 1) \quad (3)$$

where $\eta$ is random number from a uniform distribution, $0 < \eta < 1$. This amounts to replacing the first factor of the propagator with an equivalent expression, as far as quantum diffusion computations are concerned. This simple trick allows to obtain considerable code accelerations.

As for the non-conservative factor $P_2$:

$$P_2 = exp\left[\frac{-\tau(V(\mathbf{R})-V(\mathbf{R}')-2E_T)}{2}\right] \quad (4)$$

if $P_2 < 1$, a random number is generated to continue the evolution of the *i*-th walker with probability $P_2$. If this walker disappears, a flag variable $\xi(i)$ is set to zero. If $P_2 > 1$, a new walker is produced at the same position with probability $P_2 - 1$. In such case, the new walker is placed at the tail of the list of walkers. At the end of each complete update of the x's, the list of walkers is shortened by removing those with $\xi(i) = 0$.

During the simulation, $E_T$ is constantly adjusted so that the average number of walkers is approximately constant [Thijssen, 2007]:

$$E_{T_i} = E_{T_{i-1}} + \alpha ln\left(\frac{M_{i-1}}{M_i}\right) \quad (5)$$

where $\alpha$ is a small positive parameter, and $E_{T_i}$, $E_{T_{i-1}}$, $M_i$ and $M_{i-1}$ are the energy values and number of walkers at time step $i$ and $i$-1. In order to better improve the $E_T$ calculation in the code, we adjusted equation (5) by considering an average number of walkers $M_{av,i}$ (at time step $i$) in the numerator of the logarithmic term:

$$E_{T_i} = E_{T_{i-1}} + \alpha ln\left(\frac{M_{av,i}}{M_i}\right) \quad (6)$$

By doing so, the natural oscillations of the Monte Carlo method are further reduced. At the numerical steady-state, $E_T$ becomes the estimate of the energy eigenvalue.

The method always provides the energy of the ground state of the system, with a very limited number of numerical parameters to be optimized (typically: energy offset $E_T$, time step $\tau$, number of walkers $M$). We assume $\alpha$ equal to $\tau$ to reduce the number of numerical parameters.

In order to reduce energy fluctuations and statistical deviations (typical of the DMC method), we implemented a sort of restart of the walker cloud, for each point of the potential energy curve. In a first simulation, the walker population is roughly initialized in a hypercubic subset of the phase space. The walker ensemble then diffuses according to equation (1), and the energy offset is adjusted according equation (6). At the end of this simulation, a new and more realistic configuration of the walker cloud is obtained. This new configuration represents a sort of "guiding wavefunction", and it is used as a starting point for a second simulation, which then gives a more precise energy eigenvalue. In contrast with recent Monte Carlo calculations [Sarsa, 2012; Sarsa, 2014], our programs do not employ an analytic expression of a guess wavefunctions to reduce variance. Instead, it performs long imaginary time evolutions, in the course of which averages are collected.

This is the reason we name it "unbiased": it is essentially the most primitive form of the algorithm [Anderson, 1975] but with some significant improvements. The first one is the aforementioned replacement of the expression of the propagator (equation (2)) with a different one, computationally equivalent, but less computationally expensive. The second improvement consists in using the walker cloud distribution obtained in a previous calculation and the related eigenvalue estimate as starting points for a calculation with a slightly different value of the nuclear positions: in this way, the previous calculation works as an estimate of the assumption to be obtained in the new calculation.

While this technique leads to somewhat larger oscillations and fluctuations compared to the employ of analytical guesses, it has the advantage to be applicable to different confinement geometries with minimal variations, without having to change the expression of the guess wavefunction, since no such guess is used.

DMC can handle only positive distributions, so many problems arise when we want to deal with a many-electron wave function. The DMC method can address low excited states (whose wave-functions are not always positive) with distinct symmetries as well, by adding nodal surfaces that reproduces the excited state symmetry.

However, in the case of two electrons systems, an unrestricted, six-dimensional wavefunction $\psi(r_1, r_2)$ for the ground state can be calculated directly. This means that the electron correlation is accounted *exactly*, with a remarkably low analytical fatigue: no expansion of the wavefunction $\psi$ is needed, no function database needs to be included, at least in the unbiased approach.

In this work, we are dealing with such two-electron wavefunctions. Our native codes are therefore updated to simulate two-electron, confined systems thanks to the aforementioned features. All equations that govern the diffusion and branching algorithms can be changed accordingly with to necessary to abandon cartesian coordinates. As a consequence, we can apply the DMC method to confined systems, like $H_2$ and $H_3^+$, with great freedom concerning confinement geometry, by simulating the six-dimensional system directly with the additional constraint that no six-dimensional walker can cross the confining surfaces. In the following sections, we consider two test cases (cylindrical confined $H_3^+$ and crystal-field confined $H_2$) to show the versatility of the method.

## 3. Cylindrical confined $H_3^+$

In this section of the work, we show results in the case of $H_3^+$: this ion is the simplest triatomic species with only two electrons, so it is a system of considerable interest from a fundamental point of view. The confinement by cylindrically symmetrical potential surfaces represents a good model for the chemical confinement of small molecules within nanotubes, such as carbon nanotubes. A quantum computation allows us to determine the potential energy surface of these species in the confined state. The potential energy surface can then be used to make structural studies, in particular to determine the equilibrium position of the nuclei, and dynamic studies, that lead to predictions of infrared and Raman spectroscopy.

In light of the increasing importance of studies related to the confinement of hydrogen-based species, in various kind of materials specially designed to store large quantities of hydrogen (i.e., carbon nanotubes) [Coppola, 2019], it is important to undertake some first confinement studies of hydrogen-based species in nanotubes.

The cylindrical confinement of hydrogen species has been occasionally studied in the past, for example in the case of H atoms [Yurenev, 2006], and $H_2^+$ and $H_2$ molecules [Sarsa, 2014], but no result has ever been reported in the literature for the $H_3^+$ ion under such a confinement.

Since these are the first calculations for this confined system, we have considerably reduced the complexity of the problem: in particular, we are assuming that the molecule-ion is collinear, with the molecular axis placed on the nanotube axis. We also assume that the two internuclear distances, from the central nucleus to the two terminal nuclei, are equal to each other and are indicated by *d*: the full system keeps therefore a $D_{\infty h}$ symmetry (Figure 1). This limitation is assumed only to reduce the number of independent parameters: the code can treat any other symmetry without modification. The cylinder radius, *c*, is the confinement characteristic length.

This system is implemented by introducing an appropriate expression for potential energy, i.e.:

$$V(\boldsymbol{r}_1, \boldsymbol{r}_2) = -\frac{1}{|\boldsymbol{r}_1 - \boldsymbol{k}d|} - \frac{1}{|\boldsymbol{r}_1 + \boldsymbol{k}d|} - \frac{1}{|\boldsymbol{r}_2 - \boldsymbol{k}d|} - \frac{1}{|\boldsymbol{r}_2 + \boldsymbol{k}d|} - \frac{1}{r_1} - \frac{1}{r_2} \qquad (7)$$
$$+ \frac{1}{|\boldsymbol{r}_1 - \boldsymbol{r}_2|} + \frac{5}{2d}$$

where $\boldsymbol{k}$ is *z*-axis unit vector, and inserting the boundary conditions (see Figure 1)

$$\psi(\mathbf{r}_1, \mathbf{r}_2) = 0 \;\; if \;\; \begin{cases} x_1^2 + y_1^2 > c^2 & or \\ x_2^2 + y_2^2 > c^2 & \end{cases} \tag{8}$$

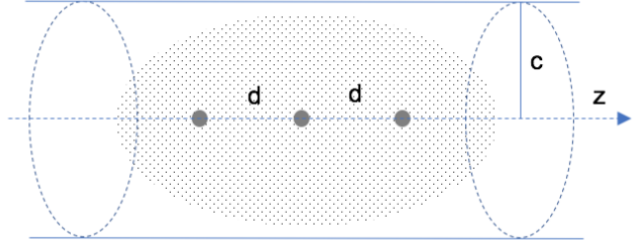

Figure 1. Collinear $H_3^+$ ($d$ is the internuclear distance) confined in a cylindrical nanotube with $D_{\infty h}$ symmetry. $c$ represents the confinement dimension. The internuclear distances, $d$, are equal.

The boundary conditions are implemented by removing walkers that enter the "forbidden" region specified by the equation (8). The results in Figures 2 and 3 show what can be obtained using the DMC method in its primitive form, without the use of test functions. Note that the excellent reproduction of the state-of-the art potential energy surface of $H_3^+$ (Figure 2) amounts to a validation of the method.

In Figure 2, the comparison with the best available results, in the case of the unconfined system, shows how a very simple DMC code algorithm is able to obtain a satisfactory prediction with small deviations from the reference points in most cases. By inspecting the figure, it is also clear that the trend of the calculated solution is satisfactory: therefore, in a future application to *ab-initio* molecular dynamics, the method may be exploited to produce a reliable interpolated potential energy surface. The potential energy curve is actually reproduced with sufficient accuracy to perform, for example, *a posteriori* calculations of vibrational states: this is not a calculation that can be done with our computational code (furthermore, it is out of the domain of the DMC method studied here), because the free collinear geometry is not stable in a vibration (the geometry of the free $H_3^+$ ion is $D_{3h}$).

At this point, thanks to the versatility of the algorithm, it is possible to introduce readily the cylindrical infinite potential barrier illustrated in Figure 1. Figure 3 shows the results of the compression of the electron cloud of the ion molecule by a cylindrical barrier with decreasing radius: the reduction of the available volume leads to a general lifting of the potential curve. The most important result is the shift towards smaller values of the equilibrium distance: this means that the molecule contracts due to the increase in the density of the electron cloud. Although similar results were reported for $H_2$ under analogous confinement conditions [Sarsa, 2014], our point here is that the more complex system $H_3^+$, with two electron, can be studied with little difficulty with our DMC formulation, by avoiding the formulation of a suitable trial function. Only equation (7) needs to be changed. A substantial increase in the binding energy (compared to dissociation) is also observed: this is another effect due to the compression of the electron cloud.

We underline that, while we have studied the collinear geometry of $H_3^+$, nothing in the code prevents the consideration of any nuclear geometry, since the code high flexibility with regard to the changing of geometry.

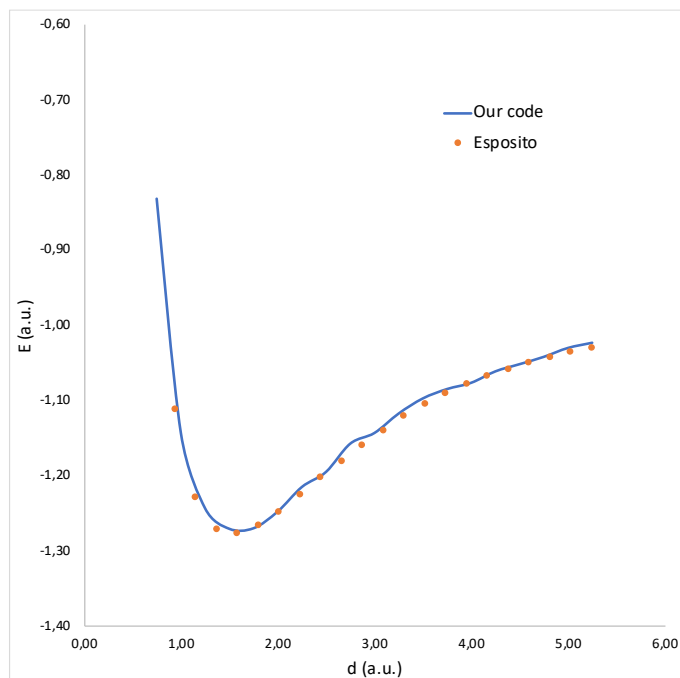

Figure 2. Free collinear $H_3^+$: validation of our potential energy curve (blue line) with state-of-the-art data (orange dots) from a recent database courtesy Dr. Esposito, CNR-ISTP). The two internuclear distances, $d$, are equal.

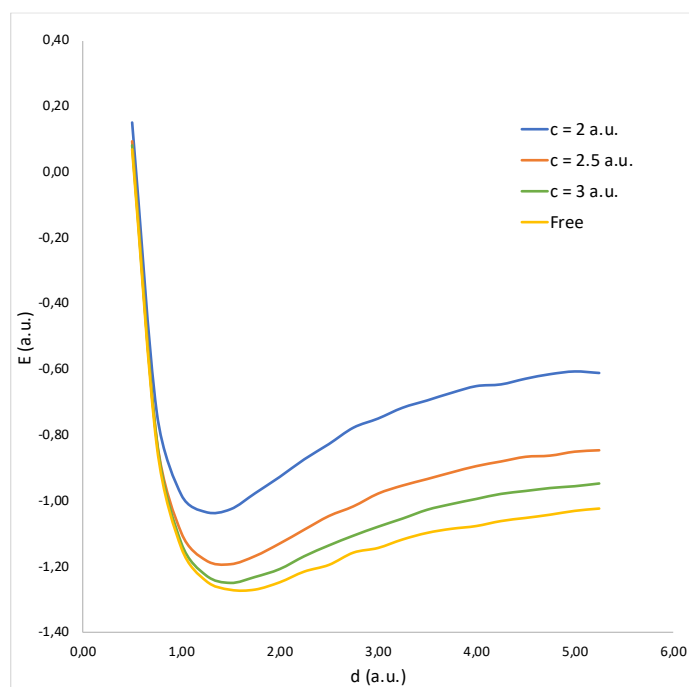

Figure 3. Collinear $H_3^+$ potential energy curves, with different confinement dimension: free and cylindrical confined $H_3^+$. $c$ represents the confinement dimension. The two internuclear distances, $d$, are equal.

## 4. Crystal-field confined $H_2$

In the previous section, we have shown the effect of a confinement by an infinite potential surface and with a geometric section: this is essentially a mathematical abstraction of more complex chemical problems. Actually, this approach (potential well confinement) was followed by the vast majority of the literature on confined hydrogen species.

With our algorithm, however, it is also possible to study a somewhat more realistic confinement mechanism, closer to the chemical reality of some systems. The mechanism, inspired by the premises of Bethe's crystalline field theory [Basolo, 1986], is based on the idea of using point electric charges in realistic geometric configurations of a crystal lattice, and which therefore produce a static potential surface. This acts on the charged species inside the molecule, in particular on the electrons.

Although this test case is not a strict geometrical confinement, it is clear that the electrostatic potentials due to the shell of negative charges acts as a confining potential well for the electrons of the enclosed molecule.

Additionally, we expect a modulation of the potential energy surface, when the species included in the crystalline field are molecular species.

As a preliminary check, we have run the code in the case of the ground state of $H_2$. At the equilibrium internuclear distance of 1.40 a.u,, with a $10^3$ walkers and 50000 times steps at $\tau = 0.002$, our code provides a ground energy of -1.171 a.u., to be compared with the exact value given by -1.1744 a.u. The accuracy can be further improved with a higher computational cost.

Figure 4 shows a sketch of the $H_2$ molecule confined in a six-negative-charge cage: the hydrogen molecule has a molecular axis directed towards two of the opposite charges of the octahedron, and its center of mass coincides with the center of mass of the octahedron. We therefore have a $D_{4h}$ symmetry.

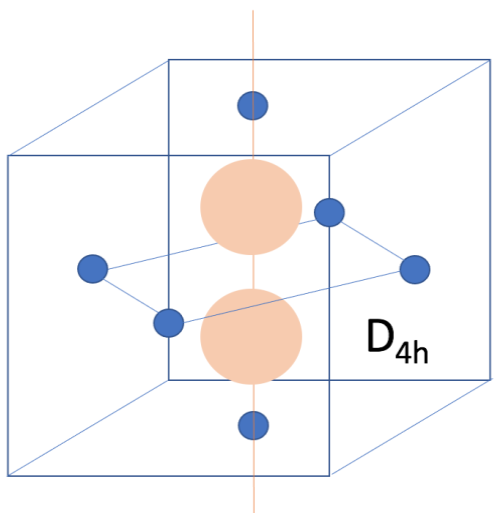

Figure 4. $H_2$ confined by a system of six-negative-point charges, with the hydrogen molecule-point charges system in a configuration of $D_{4h}$ symmetry (although the octahedron is assumed to remain geometrically perfect, the symmetry is broken by the molecule).

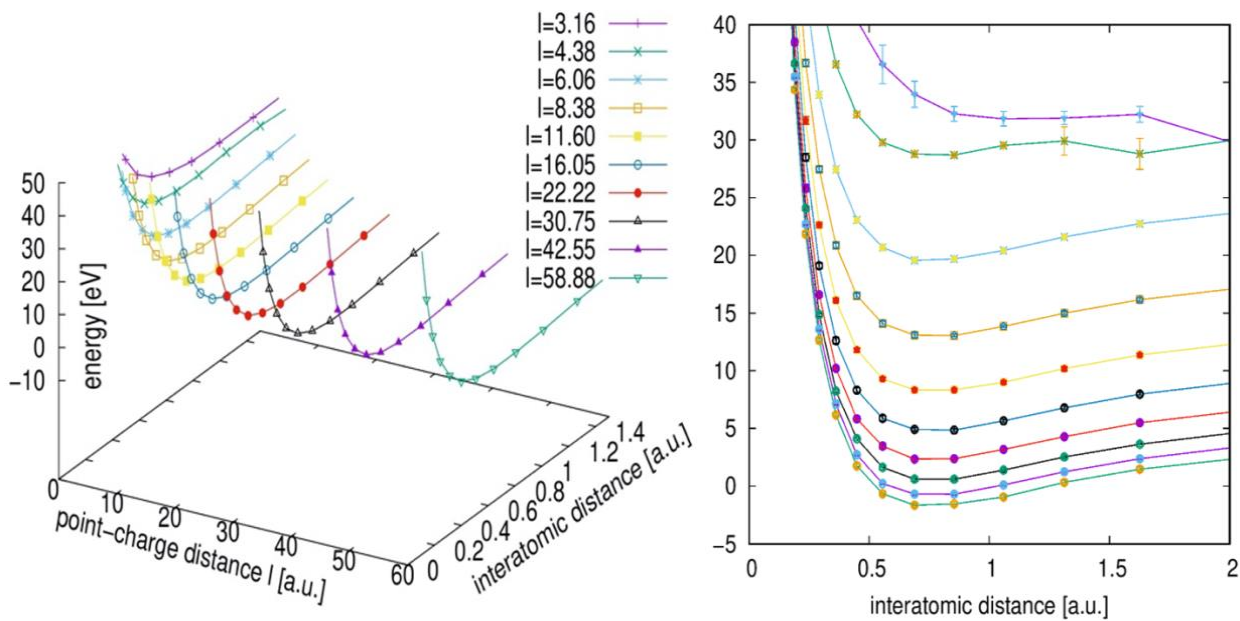

Figure 5. Potential energy curves (on the right: close-up with extended x-axis) of one $H_2$ molecule confined by a system of six-negative-point charges (equal to -$e$). The variable $l$ represents the distance of each point-charge from the center of symmetry.

In Figure 5, the energies of a confinement as from Figure 4 are reported for several distances $l$ of the point-charges from the center of the octahedron. Each point is the result of the average of ten runs; increasing $l$ brings to a larger standard deviation, which means that the methods works even better for the strongly confined systems which are the target of our studies.

Figure 5 shows some of the results that can be obtained with our code: as the point charges approach, even in this case an effect of general energy increase is observed, due to the Coulomb repulsion, and the trend of the potential energy surface is significantly modified.

Unlike the previous case study, however, a significant shift in the minimum of the potential energy curve is not observed as the confinement strength increases. In the case of very narrow confinements, however, the state becomes progressively less bound and ultimately repulsive (Figure 5). It is also probable that the minimum energy geometry of the molecule is not the one used for this preliminary study, but rather one close to a $D_{3d}$ symmetry. These aspects can be investigated with more detailed calculations.

## 5. Conclusions

In conclusion, in the context of a new emerging class of problems, our proposal of the unbiased Diffusion Monte Carlo calculation technique lies in the contrast between analytical fatigue and computational cost. The variance reduction techniques used for Monte Carlo methods require the formulation of fully correlated tentative wave functions: this formulation is not immediate in the case of confined systems, and it is not immediately transportable from one confinement geometry to another. On the other hand, the accuracy of the unbiased Monte Carlo method can be increased by accurate statistical analysis and simple solutions. Nowadays, the large availability of fast computing resources, even portable ones, might lead a researcher to favor an approach that allows to develop native, short and readable programs, and, at the same time, allows to explore different confinement geometries with minimal modifications. Even with the limits due to statistical fluctuations, such algorithm takes the correlations between electrons into full account. Furthermore, our technique, despite being in the spirit of the primitive one, includes improvements of straightforward implementation, such as the faster propagator and the method of restarting the walker cloud. Ultimately, our experience suggests that a simpler technique, employing no analytic functions of significant complexity, can be a resource in the exploration of the wide range of possibilities that has been opened up by the recent reconsideration of confined quantum systems.

**Acknowledgement**
The authors thank Dr. Fabrizio Esposito (CNR-ISTP) for the use of his databases of the free $H_3^+$ potential energy surface. Prof. Nazzareno Re, University of Chieti, is credited with the original suggestion to use point charges to produce a confining potential well.

**References**
Adamowicz, L, & Pavanello, M. (2012), *Philosophical Transactions of the Royal Society, 370*, 5001-5013.
Anderson, J. B. (1975). *The Journal of Chemical Physics, 63*(4), 1499-1503.
Anderson, J. B. (1992). *The Journal of Chemical Physics, 96*(5), 3702-3706.
Aquino, N., Campoy, G., & Montgomery, H.E. (2007) *International Journal of Quantum Chemistry*, *107,* 1548–58.
Basolo, F., & Ronald C. J. (1986) *Coordination chemistry*. Science reviews.
Belosludov, V.R., Inerbaev, T.M., Belosludov, R.V., & Kawazoe, Y. (2003) *Physical Review B, 67,* 155410.
Burrows, B.L., & Cohen, M. (2006). *International Journal of Quantum Chemistry*, *106,* 478–84.
Chuu, D.S., Hsiao, C.M., & Mei, W.N. (1992). *Physical Review B, 46*, 3898.
Colín-Rodríguez, R., & Cruz, S. A. (2010). *Journal of Physics B: Atomic, Molecular and Optical Physics*, *43*, 235102.
Connerade, J. P., Dolmatov, V. K., & Manson, S. T. (1999). *Journal of Physics B: Atomic, Molecular and Optical Physics*, *32*(14), L395.
Conroy, H. (1964). *The Journal of Chemical Physics, 40*(2), 603-604.
Coppola, C. M., et al. (2019). *Rendiconti Lincei. Scienze Fisiche e Naturali, 30*(2), 287-296.
De Groot, S. R., & Ten Seldam, C. A. (1946). *Physica, 12*(9-10), 669-682.
De Oliveira Batael, H., & Drigo Filho, E. (2018). *Theoretical Chemistry Accounts, 137*(5), 65.
Doma, S. B., El-Gammal, F. N., & Amer, A. A. (2018). *Molecular Physics, 116*(14), 1827-1833.
Foulkes, W. M. C., Mitas, L., Needs, R. J., & Rajagopal, G. (2001) *Reviews of Modern Physics*, **73(1)** 33.
Harrison, P. (2005). *Quantum Wells, Wires and Dots: Theoretical and Computational Physics of Semiconductors Nanostructures,* Wiley-Interscience, New York.
Hernández-Rojas, J., Bretón, J., & Gomez Llorente, J. M. (1996). *The Journal of chemical physics*, *104*(4), 1179-1186.
Jaskólski, W. (1996). *Physics Reports 271*, 1-66.
Jin, B., Schmider, H. L., & Wardlaw, D. (2000). *Chemical Physics Letters, 317*, 464-471.
LeSar, R., & Herschbach, D. R. (1981). *The Journal of Physical Chemistry, 85*(19), 2798-2804.
LeSar, R., & Herschbach, D. R. (1983). *The Journal of Physical Chemistry, 87*(25), 5202-5206.

Ley-Koo, E. (2018). *Revista mexicana de física*, *64*(4), 326-363.

Longo, S., Micca Longo, G., & Giordano, D. (2018). *Rendiconti Lincei. Scienze Fisiche e Naturali*, *29,* 173–177.

Micca Longo, G., Longo, S., & Giordano, D., (2015a). *Physica Scripta*, *90,* 025403.

Micca Longo, G., Longo, S., & Giordano, D., (2015b). *Plasma Sources Science and Technology*, *24,* 065019.

Micca Longo, G., Longo, S., & Giordano, D., (2015c). *Physica Scripta*, *90,* 085402.

Michels, A., De Boer, J., & Bijl, A. (1937). *Physica*, *4*(10), 981-994.

Mizus, I. I., Polyansky, O. L., McKemmish, L. K., Tennyson, J., Alijah, A., & Zobov, N. F. (2018). *Molecular Physics, 117*(13), 1663-1672.

Munjal, D., Sed, K.D., & Prasad, V. (2018). *Journal of Physics Communications, 2*, 025002.

Pang, T. (1994). *Physical Review A, 49*(3), 1709.

Pavanello, M., Tung, W. C., Leonarski, F, & Adamowicz, L. (2009), *The Journal of Chemical Physics, 130,* 074105.

Porras-Montenegro, N., & Pérez-Merchancano, S.T. (1992). *Physical Review B, 46*, 9780.

Röhse, R., Kutzelnigg, W., Jaquet, R., & Klopper, W. (1994) *The Journal of Chemical Physics, 101*, 2231-2243.

Sabin, J. R., Brandas, E., & Cruz, S. A. (Eds.) (2009). *The Theory of Quantum Confined Systems. Parts I and II. Advances in Quantum Chemistry*, Vol. 57, New York, Academic Press.

Sarsa, A., & Le Sech, C. (2012). *Journal of Physics B: Atomic, Molecular and Optical Physics*, *45*(20), 205101.

Sarsa, A., Alcaraz-Pelegrina, J. M., Le Sech, C., & Cruz, S. A. (2013). *The Journal of Physical Chemistry B, 117*(24), 7270-7276.

Sarsa A., & Le Sech C. (2014). *Study of Quantum Confinement of $H_2^+$ Ion and $H_2$ Molecule with Monte Carlo. Respective Role of the Electron and Nuclei Confinement*. In: Sen K. (eds) Electronic Structure of Quantum Confined Atoms and Molecules. Springer, Cham.

Sen, K. D., Pupyshev, V. I., & Montgomery, H. E. (2009). *Theory of confined quantum systems: part one. Special edition.* Academic Press, USA, (Adv Quant Chem 57), 25-77.

Sen, K. D., & Sen, K. D. (Eds.). (2014). *Electronic structure of quantum confined atoms and molecules*. Switzerland: Springer International Publishing.

Sommerfeld, A., & Welker, H. (1938). *Annalen der Physik, 424*(1-2), 56-65.

Soullard, J., Santamaria, R., & Cruz, S. A. (2004). *Chemical physics letters*, *391*(1-3), 187-190.

Thijssen, J. (2007). *Computational physics*. Cambridge University Press.

Yurenev, P. V., Scherbinin, A. V., & Pupyshev, V. I. (2006). *International journal of quantum chemistry 106*(10), 2201-2207.